\documentclass[aps,prb,amsmath,amssymb,showpacs,twocolumn,superscriptaddress]{revtex4}

\usepackage{dcolumn}
\usepackage{bm}
\usepackage[dvips]{graphicx}
\usepackage{subfigure}
\usepackage{pstricks}
\usepackage{epic,eepic}

\newcommand{\dnhlabel}[1]{
\setlength{\unitlength}{0.09em}
\begin{picture}(22,6)(-5,-2)
\path(4,0)(14,0)
\put(4,0){\whiten\circle{4}}
\put(-3,-4){{\scriptsize #1}}
\put(14,0){\whiten\circle{4}}
\end{picture}}

\newcommand{\dnodlabel}[1]{
\setlength{\unitlength}{0.09em}
\begin{picture}(16,9)(-9,5)
\path(0,3)(5,11.7)
\put(0,3){\whiten\circle{4}}
\put(-7,-2){{\scriptsize #1}}
\put(5,11.7){\whiten\circle{4}}
\end{picture}}

\newcommand{\dnoilabel}[1]{
\setlength{\unitlength}{0.09em}
\begin{picture}(16,6)(-7,5)
\path(5,3)(0,11.7)
\put(5,3){\whiten\circle{4}}
\put(-2,-2){{\scriptsize #1}}
\put(0,11.7){\whiten\circle{4}}
\end{picture}}

\newcommand{\dnplabel}[1]{
\setlength{\unitlength}{0.09em}
\begin{picture}(9,6)(-3,-2)
\put(4,0){\whiten\circle{4}}
\put(-2,-4){{\scriptsize #1}}
\end{picture}}

\newcommand{\dnrdi}{
\setlength{\unitlength}{0.09em}
\begin{picture}(18,10)(-3,2)
\path(0,0)(10,0)
\path(10,0)(15,8.7)
\path(15,8.7)(5,8.7)
\path(5,8.7)(0,0)
\path(5,8.7)(10,0)
\put(0,0){\whiten\circle{4}}
\put(-7,-6){{\scriptsize 1}}
\put(10,0){\whiten\circle{4}}
\put(14,-6){{\scriptsize 2}}
\put(15,8.7){\whiten\circle{4}}
\put(18,9.7){{\scriptsize 4}}
\put(5,8.7){\whiten\circle{4}}
\put(-2,9.7){{\scriptsize 3}}
\end{picture}}

\newcommand{\dnrdlabel}[1]{
\setlength{\unitlength}{0.09em}
\begin{picture}(24,10)(-9,2)
\path(0,0)(10,0)
\path(10,0)(15,8.7)
\path(15,8.7)(5,8.7)
\path(5,8.7)(0,0)
\path(5,8.7)(10,0)
\put(0,0){\whiten\circle{4}}
\put(-7,-4){{\scriptsize #1}}
\put(10,0){\whiten\circle{4}}
\put(15,8.7){\whiten\circle{4}}
\put(5,8.7){\whiten\circle{4}}
\end{picture}}

\newcommand{\dnrilabel}[1]{
\setlength{\unitlength}{0.09em}
\begin{picture}(27,10)(-9,2)
\path(5,0)(15,0)
\path(15,0)(10,8.7)
\path(10,8.7)(0,8.7)
\path(0,8.7)(5,0)
\path(5,0)(10,8.7)
\put(5,0){\whiten\circle{4}}
\put(-2,-4){{\scriptsize #1}}
\put(15,0){\whiten\circle{4}}
\put(10,8.7){\whiten\circle{4}}
\put(0,8.7){\whiten\circle{4}}
\end{picture}}

\newcommand{\dnrvlabel}[1]{
\setlength{\unitlength}{0.09em}
\begin{picture}(24,10)(-9,2)
\path(0,4.35)(5,-4.35)
\path(5,-4.35)(10,4.35)
\path(10,4.35)(5,13.05)
\path(5,13.05)(0,4.35)
\path(10,4.35)(0,4.35)
\put(0,4.35){\whiten\circle{4}}
\put(-7,4.35){{\scriptsize #1}}
\put(5,-4.35){\whiten\circle{4}}
\put(10,4.35){\whiten\circle{4}}
\put(5,13.05){\whiten\circle{4}}
\end{picture}}

\newcommand{\dntulabel}[1]{
\setlength{\unitlength}{0.09em}
\begin{picture}(22,8)(-9,1)
\path(0,0)(10,0)
\path(10,0)(5,8.7)
\path(5,8.7)(0,0)
\put(0,0){\whiten\circle{4}}
\put(-7,-4){{\scriptsize #1}}
\put(10,0){\whiten\circle{4}}
\put(5,8.7){\whiten\circle{4}}
\end{picture}}

\newcommand{\dntdlabel}[1]{
\setlength{\unitlength}{0.09em}
\begin{picture}(22,8)(-9,1)
\path(0,8,7)(10,8.7)
\path(10,8.7)(5,0)
\path(5,0)(0,8.7)
\put(0,8.7){\whiten\circle{4}}
\put(-2,-4){{\scriptsize #1}}
\put(10,8.7){\whiten\circle{4}}
\put(5,0){\whiten\circle{4}}
\end{picture}}

\begin{document}

\title{Phase diagram of a two-dimensional lattice gas model of a ramp system}

\author{No\'e G.\ Almarza}
\email{noe@iqfr.csic.es}
\affiliation{Instituto de Qu\'{\i}mica F\'{\i}sica Rocasolano, Consejo Superior
de Investigaciones Cient\'{\i}ficas (CSIC),
calle de Serrano 119, E-28006 Madrid, Spain }

\author{Jos\'e A.\ Capit\'an}
\email{jcapitan@math.uc3m.es}
\affiliation{Grupo Interdisciplinar de Sistemas Complejos (GISC), Departamento
de Matem\'aticas, Universidad Carlos III de Madrid, avenida de la Universidad 30,
E-28911 Legan\'es, Madrid, Spain}

\author{Jos\'e A.\ Cuesta}
\email{cuesta@math.uc3m.es}
\affiliation{Grupo Interdisciplinar de Sistemas Complejos (GISC), Departamento
de Matem\'aticas, Universidad Carlos III de Madrid, avenida de la Universidad 30,
E-28911 Legan\'es, Madrid, Spain}

\author{Enrique Lomba}
\email{E.Lomba@iqfr.csic.es}
\affiliation{Instituto de Qu\'{\i}mica F\'{\i}sica Rocasolano, Consejo Superior
de Investigaciones Cient\'{\i}ficas (CSIC),
calle de Serrano 119, E-28006 Madrid, Spain }

\date{\today}

\begin{abstract}
Using Monte Carlo Simulation and fundamental measure
theory we study the phase diagram of a two-dimensional
lattice gas model with a nearest neighbor hard core exclusion and a
next-to-nearest neighbors finite repulsive interaction. The model
presents two competing ranges of interaction and, in common with many
experimental systems, exhibits a low density solid phase, which melts
back to the fluid phase upon compression. The theoretical approach is
found to provide a qualitatively correct picture of the phase diagram
of our model system.
\end{abstract}


\maketitle

\section{Introduction}
The presence of liquid-liquid (LL) equilibrium in simple fluids has drawn
considerable attention in recent
years,\cite{NAT_2000_403_170,PRL_2004_92_025701,JPCM_2005_17_L293}
mainly due to its connection with the existence of certain thermodynamic,
structural and dynamic anomalies in liquid
water.\cite{JCP_1996_105_5099,PRL_2002_88_195701,JCP_2005_123_044515}
The fact that there are significant regions of the phase diagram in which an
increase of temperature at constant pressure is associated with a corresponding increase in
density, or in which diffusivity is enhanced when the system is
compressed, might be at first sight somewhat counterintuitive, and
has therefore motivated a remarkable research effort. Most of the
systems that 
exhibit this peculiar behavior are also known to present low density solid
phases (with coordination numbers ranging from 2 to 5) less dense than their
liquid counterparts. Melting upon compression is a common feature
that has to be accounted for as well. 

 In this regard, simple
models constitute a fundamental aid that can allow to identify
those essential features key to the presence of the aforementioned
anomalous behavior. Recent research has focused on two main
categories of models: orientational and isotropic. The former class
of models is constructed bearing in mind the orientational character
of the hydrogen bond interaction or the strong directional character of
the covalent bonding characteristic in systems with low density solid
phases, such as silica,\cite{PRE_2000_63_011202} 
germanium oxide\cite{JCP_1995_102_6851} or
phosphorus.\cite{NAT_2000_403_170} For this class of materials, a
series of realistic potentials have been employed in order to
characterize their anomalies via computer
simulation.\cite{JCP_1996_105_5099,PRL_2002_88_195701,JCP_2005_123_044515,NM_2003_2_739,PRB_2004_69_100101,PRE_2000_63_011202}
Useful as these studies might be, a better insight can be gained from
simpler models which can be dealt with in some cases even
analytically. Perhaps the precursor of the simple orientational models
is the Bell-Lavis two-dimensional lattice model of water,\cite{JPA_1970_3_568}
recently somewhat extended by Barbosa and
Henriques.\cite{PRE_2008_77_051204} In addition to these, the Mercedes
Benz model of water\cite{JACS_1998_120_3166}, the 3D lattice gas
model of Roberts and Debenedetti\cite{JCP_1996_105_658}, and the two-dimensional
associating lattice gas model of Henriques and Barbosa\cite{PRE_2005_71_031504,JCP_2009_130_184902} must also be
mentioned. 

The complexity of the above mentioned orientational models can be
further reduced. A weighted orientational average from these types of
interactions would lead in most cases to isotropic models with several
interaction ranges. And, since the pioneering work of Hemmer and 
Stell,\cite{PRL_1970_24_1284} it turns out that the presence of two 
competing scales or interaction ranges has been found  to lie at the
heart of the existence of multiple phase transitions in otherwise
``simple'' fluids . The ramp potential model  proposed by
Hemmer and Stell regained attention when Jagla\cite{JCP_1999_111_8980}
stressed the similarities between its behavior and the anomalous properties 
of liquid water. Since then, a good number of works have been devoted
to the continuous ramp
potential.\cite{PNAS_2005_102_16558,JPCM_2006_18_S2239,PRE_2006_74_31108,PRE_2006_73_061507,PRE_2006_74_051506,JCP_2007_126_244510} Other simple models
with competing ranges of interaction, such as 
the hard-sphere square shoulder-square well potential have also been
shown to exhibit LL equilibria.\cite{PRE_2004_69_061206} But not only
continuous models can furnish an illustrative qualitative picture of
the phase behavior and various anomalies found in water and related
systems. Isotropic lattice gas models have proven to be able to
describe the qualitative features of these systems rather
accurately. One
dimensional,\cite{JMP_1969_10_1753,JCP_2008_129_024501,hoye} two
dimensional\cite{JPCM_2004_16_8811} and three dimensional\cite{hoye}
lattice gas models have
been studied, using either mean field approaches, transfer matrix
methods and/or computer simulation. 

In this paper we will consider a two-dimensional lattice gas model
closely connected with the one studied in Ref.~\onlinecite{hoye} in three
dimensions. The model is characterized by two competing interaction
ranges (a  nearest neighbor hard core exclusion  and a finite
repulsive interaction on the next to nearest sites). This model is
strongly related to the continuum shoulder model studied in
Ref.~\onlinecite{PRE_2004_69_061206}, when the attractive interactions are
absent. Our study will focus on the reentrant melting of the low
density solid phase, using both computer simulation and Lattice Fundamental
Measure Theory
(LFMT).\cite{lafuente:2002b,lafuente:2003a,lafuente:2003b,lafuente:2004,lafuente:2005}
We will see how the theoretical approach provides a
qualitatively fairly approximate picture of our model phase diagram.

The rest of the paper is sketched as follows. A brief description of
the model is introduced in the next section. Section III is devoted to
the simulation methodology. Details on the finite size scaling
analysis of the transitions are included in Section IV. Section V
describes the LFMT as applied to this model, and finally our most significant
results and conclusions are presented in Section VI.

\section{Model}

We consider a two-dimensional model defined on a triangular lattice. 
A given site of the lattice can be either empty or occupied
by one particle. The occupation of that site excludes the
occupation of its six nearest neighbor (NN) sites. In addition 
there is a {\em repulsive} interaction
between pairs of particles located in pairs of sites that are
next to nearest neighbors (NNN).

The potential energy of an acceptable configuration is then written as:
\begin{equation}
{\cal U} = \epsilon \sum_{<ij>} n_i n_j.
\end{equation}
where $<ij>$ indicates the set of NNN pairs of sites;
the coordinates $n_k$ are equal to zero for empty sites and equal to one for
occupied sites; and $\epsilon >0$.

\begin{figure}
\includegraphics[width=38mm]{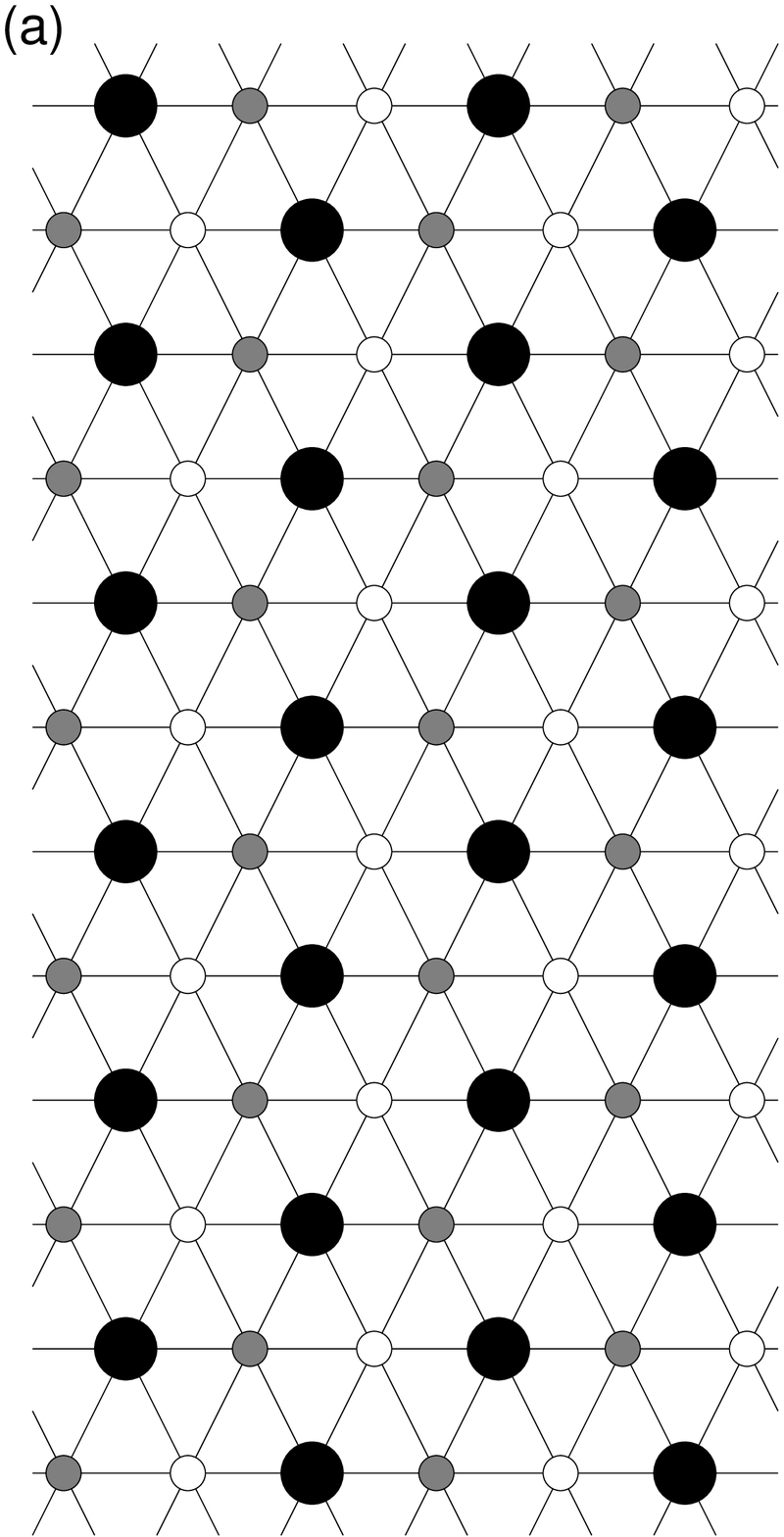}\hspace*{5mm}
\includegraphics[width=38mm]{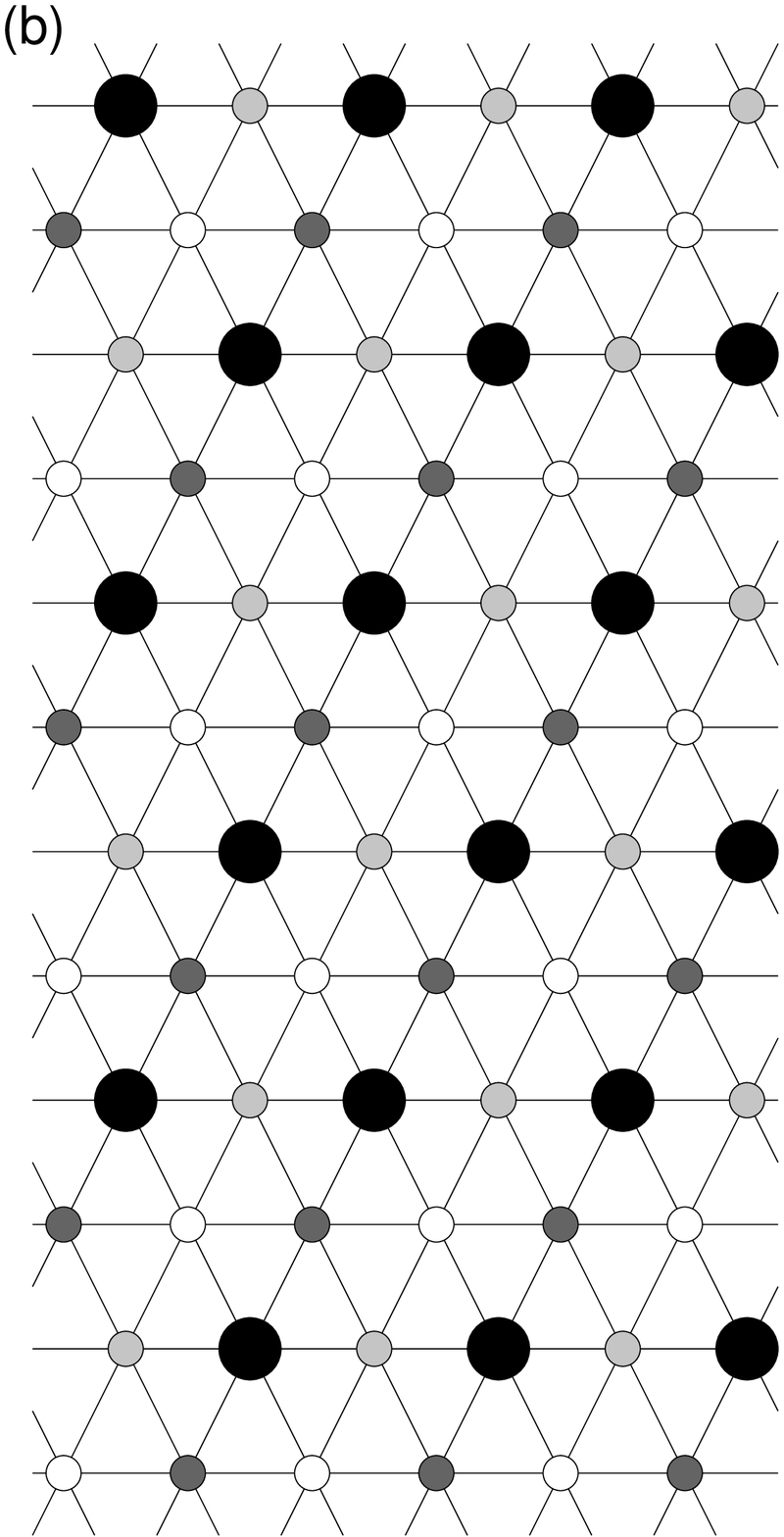}
\caption{(a) Ordered structure at high temperature
(large filled circles) and close packing.
The three sublattices are identified by circles with different shading.
Lines link nearest neighbor sites. The distance between nearest neighbors in
each sublattice is $\sqrt{3}$, i.e.\ the distance to next nearest neighbors in
the original lattice.
(b) Low density ordered structure appearing at low temperature
(large filled circles). 
The four sublattices are identified by circles with different shading.
Lines link nearest neighbor sites. The distance
between nearest neighbors in each sublattices is $2$, i.e.\ the distance
to third neighbors in the original lattice. One fourth of the sites (large circles)
of the lattice are occupied.}
\label{fig:t3}
\end{figure}

In the limit of high temperature the interactions between NNN
become negligible and the system behaves as a hard core lattice gas
with NN exclusion. Such a model is the well known
hard hexagon model, which exhibits a continuous order-disorder transition. 
The location of this transition was obtained by
Baxter,\cite{JPAMG_1980_13_L61,baxter2} and it is believed
to belong to the same universality class as the 3-state Potts model
in two dimensions.\cite{RMP_1982_54_235} At high density the system adopts an
ordered structure (that will henceforth be referred to as T3)
in which the sites occupy preferentially one of the
three sublattices of the system [see Fig.~\ref{fig:t3}(a)], with a triangular
structure. The density at close packing is $1/3$ (one third of the sites
are occupied).

In the low temperature and low pressure limit (equivalently $\epsilon\to\infty$)
another lattice hard core model is met. In this case 
an occupied site excludes the occupation of its NN and NNN.
Under these exclusion rules, at high density
we find again an ordered phase in which particles sit preferentially on
one of the four corresponding sublattices [see Fig.~\ref{fig:t3}(b)].
The close packing density is in this case $1/4$ (one fourth of
the sites are occupied), and this ordered phase will henceforth be denoted T4.
Attending to the symmetry of the order parameter and the dimensionality of the
system, one expects to find an order-disorder transition belonging to
the universality class of the 4-state Potts model in two
dimensions.\cite{RMP_1982_54_235} Theoretical analysis\cite{PRB_1984_30_5339}
and simulation results\cite{PRB_1989_39_2948,PRE_2008_78_031103} have shown
that this transition is also continuous.

As in previous work by some of the authors,\cite{hoye} we
are interested in the phase transitions of the system at intermediate
temperatures, where three phases: disordered, T4, and T3, can appear.

\section{Simulation methodology}

In order to obtain the phase diagram of the system we have made use a number
of Monte Carlo (MC) simulation techniques. At high temperature we have
used a {\em flat-histogram}
algorithm,\cite{hoye,PRE_2005_71_046132,JCP_2007_126_244510,JCP_2008_129_234504} inspired on
the Wang-Landau (WL) method,\cite{PRL_2001_86_2050,PRE_2001_64_056101}
to compute the Helmholtz energy function for all possible densities
of the systems at fixed temperature and volume.
In practice, the simulation procedure can be regarded as an extension
of the MC simulation in the grand-canonical ensemble (GCE), in which
the different number of particles are not weighted by 
a fixed chemical potential, but using a weighting function which
permits us to extend the densities entering the
sampling to an arbitrary range. Such a weighting function is not
known a priori, but it can be computed following the strategies
underlying the WL method.
Once the Helmholtz energy function is known, for a given temperature,
for all the densities and several system sizes, one can resort to
finite-size scaling strategies to locate the phase transitions.
This technique, with some modifications, has also been used to
determine the transition occurring at low temperature and moderate
density between a \emph{fluid} (F) disordered phase and a triangular
T4 phase.

In addition, we have made use of the so-called Gibbs-Duhem integration
(GDI) technique\cite{JCP_1993_98_4149,frenkel-smit} to determine the
transition between the two ordered phases and to check the consistency
of the results.

In what follows we summarize these techniques.

\subsection{Flat-histogram simulation at constant temperature}
\label{sec:flat-histogram}

The flat histogram algorithm is divided in two
parts.\cite{PRE_2005_71_046132,JCP_2007_126_244510}
The first one is devoted to find a weighting function to sample efficiently
a prefixed range of densities, at constant temperature $T$, and volume. In
this part we make use of the WL strategy. In the second part
the actual sampling of the system properties is carried out.
We will comment later about the specific details of each of these two
steps. In the application of the algorithm, two types of moves, denoted as
translation and insertion/deletion attempts, are considered.

Translational moves are carried out as follows: (i) A particle is selected at
random, and removed from its position, ${\bf R}_i$, in the system. (ii) A
trial position ${\bf R}_i^t$ (which could even be the previous one) is
selected at random with equal probability from 
those positions which are neither occupied nor excluded
by the NN interaction. (iii) The new position is accepted
with a probability given by the standard Metropolis
criterion applied to the interactions of the particle in the current
and in the trial positions, i.e.\cite{allen-tildesley,frenkel-smit}
\begin{equation}
\mathcal{A} ( {\bf R}_i^{t}|{\bf R}_i ) = \min \left\{ 1, 
\frac{ \exp \left[   - \beta {\cal U}_i({\bf R}_i^t) \right] }
     { \exp \left[  - \beta {\cal U}_i({\bf R}_i  ) \right] } \right\},
\label{acept1}
\end{equation}
where ${\cal U}_i$ is the interaction energy of particle $i$, and $\beta \equiv 1/k_B T$,
with $k_B$ being Boltzmann's constant.

In the second type of MC move, a change of the number of system particles, $N$,
is attempted. First of all it is randomly decided (with equal probabilities)
whether to increase or decrease $N$. Let us first consider the
most common case in which $N \ge 1$ for the removal attempts
and $N \le N_{\rm max}-1$ for the insertion attempts, 
$N_{\rm max}$ being the maximum number of particles to be considered.
If the number of particles is to be reduced, an occupied position is
selected at random and its particle is either removed or left according to the
acceptance criteria. If an insertion is attempted, a non-excluded position
(if there is any, otherwise the insertion attempt is directly rejected) is selected
to insert a particle, and as above the acceptance criteria are applied.
 
In order to present the acceptance probabilities of these
attempts in the context of a flat histogram procedure,
let us first write the canonical configurational partition function,
\begin{equation}
\begin{split}
Q(N,M,T) &= \frac{1}{N!} \sum_{\{{\bf R}^N\}} \exp \left[ - \beta {\cal U}({\bf R}^N) \right] \\
 &= \frac{M^N}{N!} \left\langle\exp \left[ - \beta {\cal U} \right] \right\rangle_{0},
\end{split}
\label{zn0}
\end{equation}
where $\{{\bf R}^N \}$ is the full set of $M^N$ possible configurations
of $N$ distinguishable particles over a lattice with $M$ positions. The factor
$M^N/N!$ is the contribution of the ideal lattice gas (system without interactions) and 
$\left\langle \exp \left[ - \beta {\cal U} \right] \right\rangle_{0}$ accounts for
the excess contribution to the configuration integral.

The sampling of different values of $N$
can be carried out by introducing a weight function $\omega(N)$. 
The probability of a given configuration of the system then becomes
\begin{equation}
P({\bf R}^N|M,T) \propto \omega(N) \exp  \left[-\beta {\cal U} ({\bf R}^N) \right].
\label{prn}
\end{equation}
Integrating (\ref{prn}) over all the configurations of indistinguishable particles 
for a given value of $N$ 
we get
\begin{equation}
P(N|M,T) \propto \omega(N) Q(N,M,T).
\label{pn}
\end{equation}
For the particular choice $\omega(N) = z^N$, we obtain the probability
of $N$ in the GCE:
\begin{equation}
P(N|\mu,M,T) \propto  z^N  Q(N,M,T).
\end{equation}
where $z$ is the activity, which is related with the chemical potential, $\mu$, as
$z \equiv \exp ( \beta \mu )$. 

In order to perform an effective sampling of the thermodynamics of a system
for a wide range of densities, at fixed conditions of $M$ and $T$, we can
choose a weighting function different to that defining the GC ensemble.
In practice we look for a prescription $\omega_{f}(N)$ that produces a flat
distribution $P(N|M,T)$, i.e.:
\begin{equation}
\omega_f(N) 
\propto 1/ Q(N,M,T).
\label{omegaf}
\end{equation}
For practical purposes we introduce the function ${\cal F}(N)$ defined
by $w_f(N) = N! \exp \left[ {\cal F}(N) \right]/M^N$, i.e.\
${\cal F}(N) \simeq F_{\rm ex}(N,M,T)/k_B T + K$, with $K$ 
being a constant, and $F_{\rm ex}(N,M,T)$ the excess contribution
to the Helmholtz energy function.

We can now define the acceptance probabilities of an attempted change
in the number of particles. For that we start off from the detailed
balance equation\cite{allen-tildesley,frenkel-smit,landau-binder}
\begin{equation}
\begin{split}
 P({\bf R}^{N}) & W({\bf R}^{N+1}|{\bf R}^N) {\cal A}({\bf R}^{N+1}|{\bf R}^{N}) \\
&= 
 P({\bf R}^{N+1}) W({\bf R}^{N}|{\bf R}^{N+1}) {\cal A}({\bf R}^{N}|{\bf R}^{N+1}),
\end{split}
\label{balance}
\end{equation}
where ${\bf R}^N$ and ${\bf R}^{N+1}$ are two configurations of the system with
$N$ and $N+1$ particles, respectively, which differ just in the fact that the
latter has an additional particle with respect to the former. The right-hand side
of (\ref{balance}) is the probability of moving from the configuration ${\bf R}^N$
to the configuration ${\bf R}^{N+1}$ in a given Monte Carlo step, whereas the
left-hand side corresponds to the probability of the reverse move. By $W(a|b)$
we denote the probability of choosing $a$ as trial configuration in an
insertion/deletion move, when the current configuration is $b$. Finally
${\cal A}(a|b)$ is the probability of accepting the MC move $b \rightarrow a$.

Taking into account the above described method of performing the
insertion/deletion moves we can write
\begin{eqnarray}
W({\bf R}^{N+1}|{\bf R}^N ) &=& \frac{1}{2 {\cal N}_{\rm pos}({\bf R}^{N})},
\label{wrplus} \\
W({\bf R}^{N}|{\bf R}^{N+1} ) &=& \frac{1}{2 (N+1)},
\label{wrminus}
\end{eqnarray}
where ${\cal N}_{\rm pos}({\bf R}^N)$ is the number of available positions (those
that are not excluded due to hard core interactions) in the system with
configuration ${\bf R}^N$. 
Taking into account (\ref{omegaf})--(\ref{wrminus}) we can write:
\begin{equation}
\begin{split}
\frac
{ {\cal A}({\bf R}^{N+1}| {\bf R}^{N}) }
{ {\cal A}({\bf R}^{N}| {\bf R}^{N+1}) } =& 
 \exp \left( - \beta\Delta {\cal U}_{N+1} \right)
\frac{{\cal N}_{\rm pos}({\bf R}^N)} { M} \\
& \times \exp \left[  {\cal F}(N+1) - {\cal F}(N ) \right]
\end{split}
\end{equation}
where $\Delta {\cal U}_{N+1}$ is the change of the potential energy
of the system when introducing the new particle.

We have performed two types of calculation. At high temperature
the full range of possible number of particles, $0 \le N \le M/3$, has
been sampled. In this case we have introduced transitions between the
empty and the fully occupied lattice. This is feasible because the
total number of configurations of both cases 
is known exactly, namely one for the empty lattice and three
(corresponding to the filling of each of the tree sublattices) for
the fully occupied lattice. Thus $W({\bf R}^0|{\bf R}^{M/3}) = 1/2$ and 
$W({\bf R}^0|{\bf R}^{M/3}) = 1/6$ and the corresponding acceptance
ratios can be obtained for this particular case. We will call this
scheme {\it cyclic sampling.}

At low temperatures we found difficult to sample the whole range of
densities because the WL procedure showed slow convergence. So
in order to analyze the transition between the gas and the T4 phases we
performed the simulations in the range $0\le N \le N/4$. In this case
a cyclic scheme is not feasible because at $T>0$ other
configurations different to those of the perfect T4 structure
are possible for $N=M/4$. Therefore in the insertion/deletion sampling
one directly rejects selected trials of particle deletion when $N=0$
and of particle insertion when $N=M/4$.

Technical details about how to compute the Helmholtz energy fuction
$F(N,M,T)$ using flat-histogram techniques can be found
elsewhere.\cite{PRE_2005_71_046132,JCP_2008_129_234504,JCP_2007_127_154504}
Here we will just mention the basic ideas underlying the calculation.
Simulations are divided in two parts: equilibration and sampling. 
In the equilibration part ${\cal F}(N)$ is modified
during the simulation run to push the system to visit all the values
of $N$ in the selected range. This equilibration is split in
stages, each one run until certain convergence criteria are
satisfied. As the stages go on the changes in 
${\cal F}(N)$ are smaller. At the end of the equilibration one expects
to have an appropriate estimation of ${\cal F}(N)$. Once
the equilibration part is finished, the resulting function
${\cal F}(N)$ is kept fixed and the sampling part of the simulation
starts. During this part one computes the probability of each value of
$N$ (from which a refined result for the Helmholtz energy function
$F(N,M,T)$ can be obtained) and
different properties of the system, such as energy, energy fluctuation,
order parameter, etc.  The sampling part is divided into blocks in order
to estimate error bars.

\subsection{Gibbs-Duhem integration}

As in previous works\cite{JCP_2007_126_244510,hoye} with systems exhibiting
a similar phase behavior, we have employed GDI\cite{JCP_1993_98_4149,frenkel-smit}
in the computation of the phase diagram. In the present case GDI was used
to determine the T3--T4 transition and to check the consistency of the flat histogram
MC calculations.

The analysis of the Helmholtz energy function in the limit $T\rightarrow 0$ leads to a value
of $\mu/\epsilon=12$ for the T3--T4 transition. After a number of short calculations we
conclude that such a value hardly varies for low temperatures. 

GDI goes as follows. For fixed volume (M) systems, the changes in the grand
potential can be written as
\begin{equation}
d ( - \beta p M) = {\cal U} d \beta - N d (\beta \mu).
\end{equation}
At given $\beta$ and $\mu$ phase equilibrium exists if
the pressure, $p$, is equal in both phases.
Now if one changes, for instance, $\beta$, the change in $\beta \mu$
to keep phase equilibrium is given by:
\begin{equation}
d (\beta \mu) =  \frac{ \Delta \bar{u}}{\Delta \rho}d \beta,
\label{dbetamu}
\end{equation}
where $\bar{u} \equiv {\cal U}/M$, $\rho \equiv N/M$ is the density, and $\Delta X$ denotes the
the difference of the values of the property $X$ in the two phases.
Equation~(\ref{dbetamu}), or some variants of it, can be used to
build up numerical integration schemes to compute
the phase equilibrium of discontinuous transitions.
In practice we have performed GDI using two different integration
schemes. At low temperature we have used
\begin{equation}
d \mu   = \left[ \mu - \frac{\Delta {\bar u}}{\Delta \rho} \right] d T,
\label{gdi1}
\end{equation}
whereas for the intermediate temperatures, like those at which the
T3--T4 line is expected to meet the F--T4 line transforming it into a
T3--F line, we employed
\begin{equation}
d \beta = \frac{ \Delta \rho}{\Delta {\bar u}} d (\beta \mu).
\label{gdi2}
\end{equation}

\section{Analysis of the phase transitions}

In order to locate the order-disorder phase transitions at high
and low temperatures we can use the results for
$F(N,M,T)$ to obtain the probabilities in the GC ensemble,
\begin{equation}
P(N|\mu,M,T) \propto \exp \left\{ -\beta F(N,M,T)  + \beta  \mu N  \right\}.
\end{equation}
Then we search for the value of the chemical potential, $\mu_c(M,T)$, that maximizes
the density fluctuations. For this conditions we compute the average density,
$\rho_c(M,T)$, and the momenta of the distribution of densities
$m_n(M,T) = \langle(\delta \rho)^n\rangle_{M,T}$,
with $\delta \rho = \rho - \langle \rho_c(M,T) \rangle$, for $n$ = 2, 3, and 4.

According to the definition of $\mu_c(M,T)$, we must have $m_3(M,T) =0$.
The system size dependence of $m_2(M,T)$ and the ratio $g_4(M,T) \equiv
m_4(M,T)/[m_2(M,T)]^2$, allow us to characterize the (possible) phase transition.

In principle, given the symmetry of the model, one expects that at high temperature
the phase transition will be continuous and belong to the universality class of
the 3-state Potts model in two dimensions,\cite{RMP_1982_54_235}
whereas at low temperature the order-disorder (F--T4) transition is expected to
lie in the universality class of the 4-state Potts model in two dimensions.

The scaling behavior that standard finite size scaling (FSS)
predicts\cite{landau-binder,JPAMG_1995_28_6289,PRE_1995_62_602} goes as follows:
\begin{eqnarray}
\mu_c(L,T) &\simeq & \mu_c(T) + a_{\mu} L^{-1/\nu},
\label{scal1} \\
\rho_c(L,T) &\simeq & \rho_c(T) + a_{\rho} L^{1/\nu-d},
\label{scal2} \\
m_2(L,T) &\simeq & a_{m2} L^{\alpha/\nu-d},
\label{scal3}
\end{eqnarray}
where we have used $L$ (related with $M$ by $M=2L^2$) as the system length.
These scaling laws are expected to be satisfied for large values of $L$.
In these equations $d$ is the dimension of the lattice ($d=2$), and
$\nu$ and $\alpha$ are critical exponents, which are expected to take
the values\cite{RMP_1982_54_235} $\nu=5/6$, $\alpha=1/3$ for the F--T3
continuous transition, and $\nu=2/3$, $\alpha=2/3$ for the F--T4 continuous
transition. 
  
At intermediate temperatures the nature of the transitions can change, and
eventually become first order. This fact can be studied by analyzing
the behavior of $g_4(L,T)$ with the system size. In general, 
for discontinuous transitions the value of
$g_4$,  goes to 1 as $L \rightarrow \infty$, signaling the presence of
two well defined narrow peaks in the distribution probability of the
density.

\section{Theoretical approach}
\label{sec:LFMT}

We have performed a theoretical analysis of this model using
LFMT.\cite{lafuente:2002b,lafuente:2003a,lafuente:2003b,lafuente:2004,lafuente:2005}
This theory is the lattice counterpart of Rosenfeld's Fundamental Measure
Theory\cite{Rosenfeld:1989} (FMT), and its construction is based on the
approach through zero-dimensional (0d) cavities and dimensional crossovers
of Tarazona and Rosenfeld.\cite{tarazona:1997} In short this theory amounts
to computing the exact functional for a certain set of small graphs (the
0d cavities) and then build the simplest functional for the
whole lattice which provides the exact result for density profiles that are
zero everywhere in the lattice except in a 0d cavity.
In essence this theory is the grand-canonical functional version of
Kikuchi's cluster variation method\cite{kikuchi:1951} in Morita's
formulation.\cite{morita:1994}

For short-range interacting lattice gases, the construction of a LFMT
density functional is particularly simple.\cite{lafuente:2004} Here we
will explain in detail how to apply it to the concrete model we are
studying. For a full account of the theory in all its details and with
all its properties the reader is referred to
Refs.~\onlinecite{lafuente:2004} and \onlinecite{lafuente:2005}.

The starting point of the theory is the choice of a set of so-called
\emph{maximal 0d cavities}. Zero-dimensional cavities are subgraphs of
the lattice such that every two particles placed on them necessarily
interact. They are ``maximal'' if adding a new node to the graph breaks
down this 0d requirement. If the interaction is purely hard-core
exclusion, every 0d cavity can hold just a single particle. If on top
of that there is a soft interaction, then more than one particle can
be present in a 0d cavity.
For the particular model we are considering, in a triangular lattice
$\mathcal{L}$, the set of maximal cavities is given by
\begin{equation}
\mathcal{W}_4=\bigcup\limits_{{\bf r}\in\mathcal{L}}\mathcal{W}_4({\bf r}),
\qquad
\mathcal{W}_4({\bf r}) = \bigg\{\dnrdlabel{{\bf r}},\ \dnrilabel{{\bf r}},\
\dnrvlabel{{\bf r}} \bigg\}.
\label{eq:maxcavities}
\end{equation}
The label ${\bf r}$ in the graphs denotes the position of the node beside
it (the remaining nodes of the graph are labelled accordingly). Notice
that cavities placed at different positions in the triangular lattice are
considered different.

The second step is to complete this set by closing it with respect to
non-empty intersections, i.e.\ if two overlapping cavities are in the set,
so must be their
intersection. The full set of cavities resulting from this operation,
$\mathcal{W}$, can be described as
\begin{equation}
\mathcal{W}= \bigcup\limits_{i=1}^4\mathcal{W}_i,
\end{equation}
where $\mathcal{W}_4$ is given by (\ref{eq:maxcavities}), and similarly
are defined $\mathcal{W}_i$, $i=1,2,3$, with
\begin{eqnarray}
\mathcal{W}_3({\bf r}) &=& \big\{\dntulabel{{\bf r}},\ \dntdlabel{{\bf r}} \big\}, \\
\mathcal{W}_2({\bf r}) &=& \big\{\dnhlabel{{\bf r}},\ \dnodlabel{{\bf r}},\
\dnoilabel{{\bf r}} \big\}, \\
\mathcal{W}_1({\bf r}) &=& \big\{\dnplabel{{\bf r}} \big\}.
\end{eqnarray}
The density-functional form that LFMT prescribes for this lattice gas
is then $F[\rho]=F_{\rm id}[\rho]+F_{\rm ex}[\rho]$, where
\begin{equation}
\beta F_{\rm id}[\rho]=\sum_{{\bf r}\in\mathcal{L}}\rho({\bf r})\left[\ln\rho({\bf r})-1
\right]
\label{eq:Fid}
\end{equation}
and\cite{lafuente:2004,lafuente:2005}
\begin{equation}
\beta F_{\rm ex}[\rho]=
\sum_{C\in\mathcal{W}}\left[-\mu_{\mathcal{W}}(C,\mathcal{L})\right]\Phi_0(C),
\end{equation}
with $\mu_{\mathcal{W}}(C,\mathcal{L})$ a combinatorial
object known as the M\"obius function of the
set $\mathcal{W}\cup\{\mathcal{L}\}$,\cite{lafuente:2005, aigner:1979}
and $\Phi_0(C)$ the excess free-energy density functional of the system when
constrained to be within the cavity $C$. $\Phi_0(C)$ is therefore a function
of the density profile $\rho({\bf r})$ at the nodes of $C$ alone. We will 
return to its calculation in brief.

The M\"obius coefficients $\mu_{\mathcal{W}}(C,\mathcal{L})$ satisfy the recursion
\begin{eqnarray}
\mu_{\mathcal{W}}(C,\mathcal{L}) &=& -1-\sum_{C\subsetneq C'\in\mathcal{W}}
\mu_{\mathcal{W}}(C',\mathcal{L}),
\label{eq:recursion}
\end{eqnarray}
with which it can be obtained for any cavity $C$ of the set $\mathcal{W}$.
It turns out that every cavity $C\in\mathcal{W}_i$ is contained in the same number
of cavities $C'\in\mathcal{W}_j$ with $i<j$. We shall denote this number
$M_{ij}$. Then $\mu_{\mathcal{W}}(C,\mathcal{L})$ is the same for all
cavities of the same set $\mathcal{W}_i$. So by denoting
\begin{equation}
\mu_{\mathcal{W}}(C,\mathcal{L})= m_i,
\qquad \forall C\in\mathcal{W}_i,
\end{equation}
recursion (\ref{eq:recursion}) becomes
\begin{equation}
m_i=-1-\sum_{j>i}M_{ij}m_j.
\end{equation}
It is easy to find that matrix $M=(M_{ij})$ is
\begin{equation}
M=
\begin{pmatrix}
0 & 6 & 6 & 12 \\
0 & 0 & 2 & 5 \\
0 & 0 & 0 & 3 \\
0 & 0 & 0 & 0
\end{pmatrix}.
\end{equation}
On the other hand, it follows from (\ref{eq:recursion}) that $m_4=-1$. This determines the
remaining coefficients as $m_3=2$, $m_2=0$, and $m_1=-1$, and therefore the functional as
\begin{equation}
\beta F_{\rm ex}[\rho]=\,\sum_{C\in\mathcal{W}_4}\Phi_0(C)
-2\sum_{C\in\mathcal{W}_3}\Phi_0(C)+\sum_{C\in\mathcal{W}_1}\Phi_0(C) .
\label{eq:Fex}
\end{equation}

Let us now compute the functions $\Phi_0(C)$ for all $C\in\mathcal{W}$. Actually it is
enough to obtain this function only for the maximal cavities (those of $\mathcal{W}_4$),
because any other cavity $C$ must be ---by construction--- a subgraph of one of the maximal
cavities, and therefore $\Phi_0(C)$ can just be obtained by setting $\rho({\bf r})=0$ at
the nodes of the maximal cavity which do not belong to $C$. On the other hand, by
symmetry the functional dependence of $\Phi_0(C)$ on the densities at the nodes of $C$
will be the same for all the maximal cavities of this model. In other words, the only
function we need to obtain is 
\begin{equation}
\Phi_0(\boldsymbol\rho)=\Phi_0\bigg(\,\,\dnrdi\,\,\,\,\bigg),
\end{equation}
where the cavity nodes are labelled generically and should be appropriately replaced by
the nodes of the corresponding cavity. Hence this is a function of
$\boldsymbol\rho=(\rho_1,\rho_2,\rho_3,\rho_4)$.

We start off by writing down the grand-canonical partition function for such a cavity,
namely
\begin{equation}
\Xi = 1+z+\kappa z_1z_4,
\label{eq:Xi}
\end{equation}
where $\kappa\equiv e^{-\beta\epsilon}$ (therefore we have $0<\kappa<1$ for any
$\epsilon>0$),
$z=\sum_i z_i$ denotes the total activity, and $z_i=e^{\beta[\mu-V_{\rm ext}(i)]}$,
$V_{\rm ext}(i)$ representing any external field acting on node $i$. In obtaining
(\ref{eq:Xi}) we have made use of the fact that the cavity can accommodate at most two
particles, and this only if they occupy nodes $1$ and $4$. In that case they interact
through the soft potential. From (\ref{eq:Xi}) we can obtain the densities as
$\rho_i=(z_i/\Xi)\partial\Xi/\partial z_i$, and the correlation between nodes $1$ and
$4$ as $\rho_{14}=(z_1z_4/\Xi)\partial^2\Xi/\partial z_1\partial z_4$. This yields
\begin{eqnarray}
\Xi\rho_{1(4)} &=& z_{1(4)}+\kappa z_1z_4,\label{eq:rho14} \\
\Xi\rho_{2(3)} &=& z_{2(3)}, \label{eq:rho23}\\
\Xi\rho_{14} &=& \kappa z_1z_4.
\label{eq:corr}
\end{eqnarray}
Adding the equations for $\rho_i$ up we obtain
\begin{equation}
\Xi\rho=z+2\kappa z_1z_4=\Xi-1+\Xi\rho_{14},
\end{equation}
where $\rho=\sum_i \rho_i$. Then
\begin{equation}
\frac{1}{\Xi}=1-\rho+\rho_{14}.
\label{eq:oXi}
\end{equation}
On the other hand, from Eqs.~(\ref{eq:rho14})--(\ref{eq:corr}) it follows that
$z_{1(4)}=\Xi(\rho_{1(4)}-\rho_{14})$, hence substituting this expressions in
(\ref{eq:corr}) and using (\ref{eq:oXi}) we obtain
\begin{equation}
\kappa (\rho_1-\rho_{14})(\rho_4-\rho_{14})=\rho_{14}(1-\rho+\rho_{14}).
\end{equation}
The solution to this second-order equation for $\rho_{14}$ is
\begin{equation}
\begin{split}
\rho_{14}(\boldsymbol\rho)= &\,\frac{1}{2(1-\kappa)}\Big\{-1+\rho-\kappa(\rho_1+\rho_4) \\
 & \left. +\sqrt{[1-\rho+\kappa(
\rho_1+\rho_4)]^2+4\kappa(1-\kappa)\rho_1\rho_4}\right\},
\end{split}
\label{eq:rho14clean}
\end{equation}
Finally we get $\Phi_0$ through a Legendre transform, i.e.\
\begin{equation}
\begin{split}
\Phi_0(\boldsymbol\rho) =&\, \sum_{i=1}^4 \rho_i\ln(z_i/\rho_i)
-\ln\Xi \\
=&\, \rho+(1-\rho)\ln(1-\rho+\rho_{14}) \\
&+\sum_{i=1,4}\rho_i\ln\left(1-\frac{\rho_{14}}{\rho_i}\right).
\end{split}
\label{eq:Phi0}
\end{equation}
As explained above, we obtain the $\Phi_0$ for the non-maximal cavities by setting
$\rho_i=0$ at the corresponding nodes. This leads to
\begin{equation}
\Phi_0\big(C)=\rho+(1-\rho)\ln(1-\rho), \qquad \rho=\sum_{i\in C}\rho_i,
\label{eq:PhiC}
\end{equation}
for all $C\in \mathcal{W}_1\cup \mathcal{W}_2\cup \mathcal{W}_3$.

Equations~(\ref{eq:Fid}), (\ref{eq:Fex}) and (\ref{eq:rho14clean})--(\ref{eq:PhiC})
complete the prescription for the density functional.

\section{Results}

\subsection{Monte Carlo Simulation results}

The simulations were carried out on rectangular lattices of different sizes,
built by replicating in both directions those shown
in Fig.~\ref{fig:t3} ---which depict
rectangular lattices containing $M=2\times L \times L$ sites (with $L=6$).

\subsubsection{F--T3 transition}

For the location of the F--T3 transition we used the WL cyclic sampling
described in Sec.~\ref{sec:flat-histogram},
for values of $\beta\epsilon =$
0.00, 0.10, 0.20, 0.30, 0.40, 0.50, and 0.63. For each of these values
simulations were carried out for ten system sizes:
$L=$ 12, 18, 24, 30, 36, 42, 48, 54, and 60. We computed the pseudo-critical
quantities $\beta \mu_c(L,T)$, $\rho_c(L,T)$, etc, and extrapolated these data
to the thermodynamic limit. To take into account possible deviations
from the scaling laws due to the relatively small system sizes, we use the
{\it ad-hoc} fitting
\begin{equation}
X_c(L,T) = X_c(T) + \sum_{k=1}^m a_{xk} L^{-kb_x},
\label{fit-ft3}
\end{equation}
where $X_c$ represent some physical property at the transition point,
and $b_x$ the critical exponent appearing in its corresponding
scaling law [c.f.~Eqs.~(\ref{scal1})--(\ref{scal2})]. The number $m$ is chosen
to be either one or two, according to a chi-square test.\cite{NumericalRecipes}

We have found that for most values of $\beta\epsilon$ the pseudo-critical chemical
potentials $\mu_{c}(L,T)$ can be fitted for the whole set of system sizes using
a second-degree polynomial [$m=2$ in Eq.~(\ref{fit-ft3})]. In the particular
case of $\beta\epsilon = 0.63$ the smallest system sizes ($L=12,L=18)$ were discarded
due to the interference with the F--T4 transitions.

\begin{table}
\begin{center}
\begin{ruledtabular}
\begin{tabular} {lll}
\multicolumn{1}{c}{$\beta\epsilon$} &
\multicolumn{1}{c}{$\beta \mu_c $} &
\multicolumn{1}{c}{$\eta_c $}  \\
\hline
 0.00   &  2.406(2)      &  0.8279(9)  \\
 0.10   &  3.152(2)      &  0.8430(10)  \\
 0.20   &  3.927(2)      &  0.8561(11)  \\
 0.30   &  4.728(3)      &  0.8671(6)  \\
 0.40   &  5.546(9)      &  0.8765(20)  \\
 0.50   &  6.393(3)      &  0.8840(6)  \\
 0.63   &  7.510(11)     &  0.892(4) \\
\end{tabular}
\end{ruledtabular}
\caption{Computed points for the F--T3 transition. Error bars are given in
brackets, in units of the last figure of the property, and correspond to a
confidence level of about 95\%.}
\label{table-ft3}
\end{center}
\end{table}

\begin{figure}
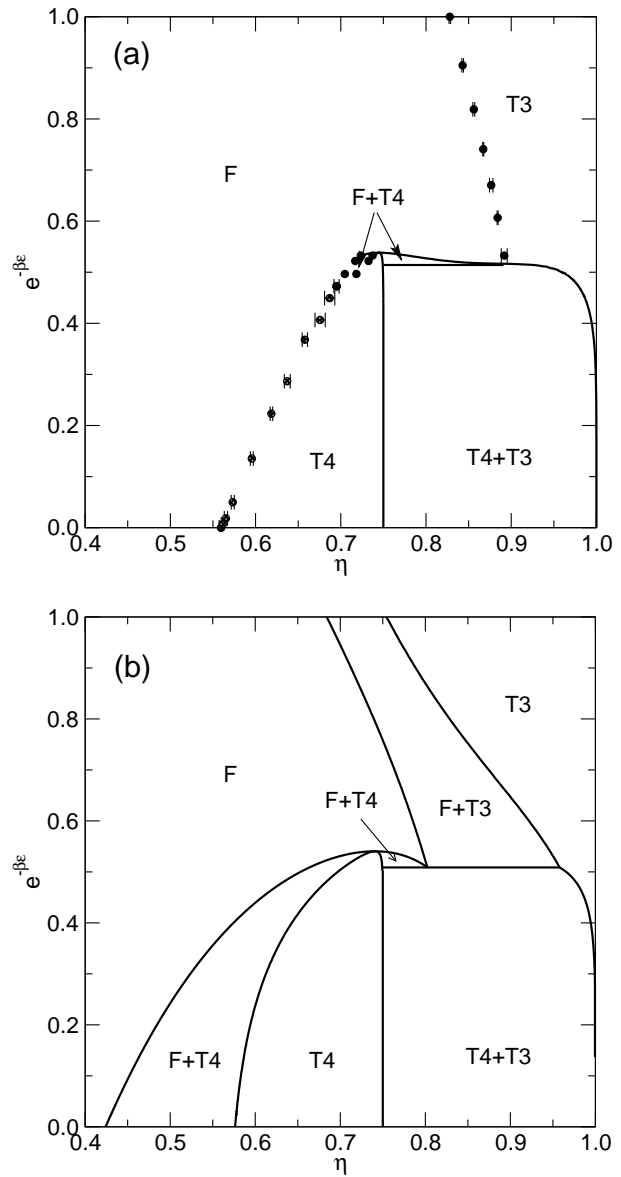

\subfigure{\includegraphics[width=80mm,clip=]{T-eta-simu.eps}
\label{T-eta-sim}}\\
\subfigure{\includegraphics[width=80mm,clip=]{T-eta.eps}
\label{T-eta-fmt}}
\caption{Temperature-density phase diagram from Monte Carlo simulations (a) and
LFMT (b). The three phases are labeled F for the fluid, and T3 and T4 for the two
solids. Coexistence regions are marked with two labels.}
\label{T-eta}
\end{figure}

\begin{figure}
\includegraphics[width=80mm,clip=]{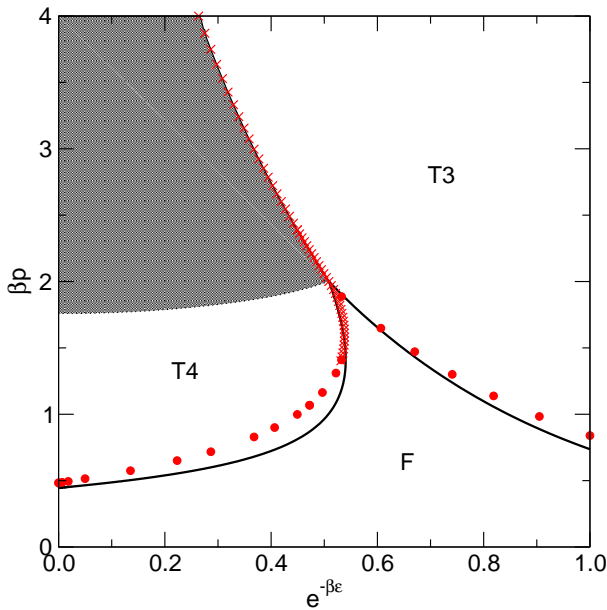}
\caption{Pressure-temperature phase diagram. The vertical axis represents the
pressure divided by $k_B T$ ($\beta p$) and the horizontal axis represents the
temperature in the form $e^{-\beta\epsilon}$, with $\epsilon>0$ the soft
repulsion between NNN sites. Symbols are the Monte Carlo simulations and lines
represent the predictions of the LFMT. The shaded region is a set of values
that cannot be reached because the pressure jumps discontinuously at the
T3--T4 transition.}
\label{bP-T}
\end{figure}

\begin{figure}
\includegraphics[width=80mm,clip=]{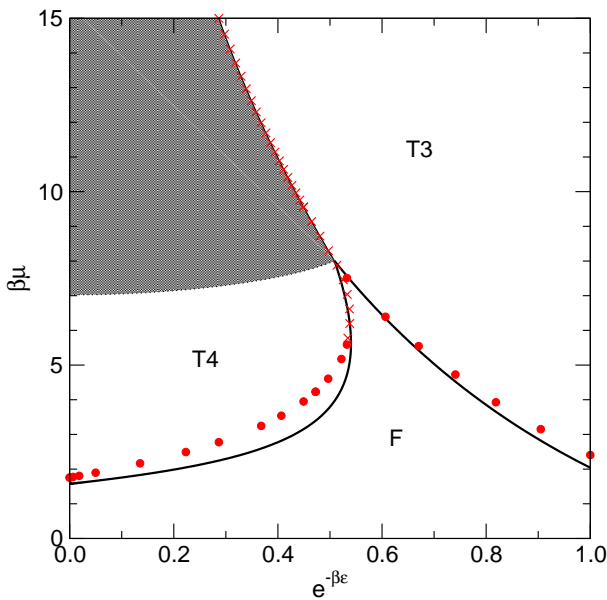}
\caption{Same as Fig.~\ref{bP-T} but for the chemical potential $\mu$.
The shaded region is a set of values that cannot be reached because
the chemical potential jumps discontinuously at the T3--T4 transition.}
\label{bmu-T}
\end{figure}

The results for the F--T3 transition are collected in Table~\ref{table-ft3} (notice
that the densities are expressed in terms of packing fractions, i.e.\ $\eta = 3 
\langle N \rangle /M$),
and plotted in Figs.~\ref{T-eta-sim}, \ref{bP-T}, and \ref{bmu-T}. The particular
case $\beta\epsilon =0 $ corresponds to the so-called hard hexagon model, whose critical
properties are known exactly.\cite{JPAMG_1980_13_L61,baxter2} The exact values are
$\beta \mu_c = \log \left[(11 + 5 \sqrt{5} )/2 \right] \approx 2.4061 $, 
and $\eta_c = 3 \rho_c = 3 ( 5 - \sqrt{5} )/10 \approx 0.8292$. A comparison with
the extrapolations given in Table~\ref{table-ft3} suggests that the latter are
quite accurate ---though not perfect--- in the estimation of the exact values.

The results for $\eta_c(\beta\epsilon)$ and $\beta \mu_c(\beta\epsilon)$
are well represented
by the fits
\begin{eqnarray}
\eta_{\rm F-T3}(\beta\epsilon) &=& 0.8280 + 0.1582 \beta\epsilon
- 0.0922 (\beta\epsilon)^2, \\
\beta \mu_{\rm F-T3}(\beta\epsilon) &=& 2.4056 + 7.3074 \beta\epsilon
+ 1.6079(\beta\epsilon)^2 \nonumber \\
  &&-0.5478 (\beta\epsilon)^3 .
\end{eqnarray} 

\subsubsection{F--T4 transition}

As in the previous case, we have selected a number of representative temperatures and
performed simulations for several system sizes. In most cases we considered
the same sizes ($L$=12, 18, $\cdots$, 60) as for the F--T3 transition. In
addition to $\beta \mu_c(L,T)$ and $\rho_c(L,T)$ we also payed attention to the 
quantities $m_2(L,T)$ and $g_4(L,T)$ because this transition appears to be discontinuous
in some cases.

The precise location of the {\em multi-critical} point ---where the transition
changes from continuous to discontinuous--- is a hard task due to the
first order transition being quite weak. In order to make an approximate
estimation we focused on the changes of $g_4(L,T)$ with $L$ and found that
the change of character of the transition occurs at about
$\beta\epsilon \approx 0.75 \pm 0.05 $. Surprisingly, it is precisely at this
temperature that the simulation results are well represented by the scaling
law (\ref{scal3}).

For $\beta\epsilon > 0.75$ we have estimated the critical properties of the F--T4 line 
in the thermodynamic limit using the same strategy as for the F--T3 case,
employing the critical exponents of the 4-state Potts model in two dimensions.
The results are gathered in Table~\ref{table-ft4}. 
The results at low temperature show a good agreement 
with those reported by Zhang and Deng \cite{PRE_2008_78_031103}: $\beta \mu_c = 1.75682(2)$, 
and $ \eta_c = 0.540(12)$.

\begin{table}
\begin{center}
\begin{ruledtabular}
\begin{tabular} {lll}
\multicolumn{1}{c}{$\beta\epsilon$} &
\multicolumn{1}{c}{$\beta \mu_c $} &
\multicolumn{1}{c}{$\eta_c $}  \\
\hline
 1000     &  1.756(1)      &  0.560(2)  \\
 10       &  1.756(2)      &  0.560(2)  \\
 5        &  1.774(1)      &  0.562(2)  \\
 4        &  1.806(1)      &  0.565(2)  \\
 3        &  1.897(1)      &  0.573(2)  \\
 2        &  2.165(1)      &  0.596(2)  \\
 1.5      &  2.491(2)      &  0.619(2)  \\
 1.25     &  2.777(3)      &  0.637(4)  \\
 1.0      &  3.246(4)      &  0.658(4)  \\
 0.9      &  3.541(3)      &  0.676(5)  \\
 0.8      &  3.950(2)      &  0.688(4)  \\
 0.75     &  4.230(2)      &  0.695(3) 
\end{tabular}
\end{ruledtabular}
\caption{Computed points for the continuous F--T4 transition. Error bars are given
in brackets, in units of the last figure of the property, and correspond to a
confidence level of about 95\%.}
\label{table-ft4}
\end{center}
\end{table}

For  $\beta\epsilon  < 0.75$ the plot of the chemical potential as a function
of $\eta$ shows a loop for the different system sizes
considered. This is a signature of a first order phase transition. The change in
density is quite small, and the transition is rather weak. Taking this into
account we have estimated the location of the transitions by fitting the
results for each simulated temperature to equations of the form (\ref{fit-ft3}),
with $b_x=1$, $m=2$. The properties considered were $\mu_c(L,T)$, $\eta_c(L,T)$,
and $\Delta \eta_c(L,T) = \sqrt{m_2(L,T)}$. Notice that, in general,
for discontinuous transitions 
$\lim_{L \rightarrow \infty} m_2(L,T) \ne 0 $. Thus, in the thermodynamic
limit, the packing fractions of the two coexisting phases are then obtained as
\begin{eqnarray}
\eta_F(T) & = & \eta_c(T) - \Delta \eta_c(T) \\
\eta_{T4}(T) & = & \eta_c(T) + \Delta \eta_c(T)
\end{eqnarray}

The results for the discontinuous F-T4 transition are collected in Table~\ref{table-dft4}.

\begin{table}
\begin{center}
\begin{ruledtabular}
\begin{tabular} {lllll}
$\beta\epsilon$               &  0.70     & 0.65     & 0.63  \\ \hline
$\beta \mu_{F-T4} $  & 4.61(2)  & 5.18(2)  & 5.59(4)   \\
$\eta_F $         &  0.705(2) & 0.717(2) & 0.724(2) \\
$\eta_{T4} $      &  0.718(2) & 0.733(2) & 0.737(2)  \\
\end{tabular}
\end{ruledtabular}
\caption{Computed points for the discontinuous F-T4 transition. Error bars are given
between parentheses, in units of
the last figure of the property and correspond to a confidence level of about 95 \%}
\label{table-dft4}
\end{center}
\end{table}

\subsection{Gibbs-Duhem Integration}

After a number of tests we analyzed the discontinuous T3--T4 transitions and its
continuation as F--T4 transition using GDI with lattices of size $L=120$. This
relatively large size was chosen since the end of this line at low values of
$\beta \mu$ corresponds to the equilibrium between a low density fluid and the
T4 phase. This transition is relatively weak and because of this we found that
for smaller values of $L$ one of the subsystems often undergoes a phase transition
and both subsystems end up in the same phase, with the corresponding breakdown
of the GDI scheme. The use of large simulation boxes then makes possible us to reach values
of $\beta \mu$ that allowed us to check the consistency between GDI and WL simulations.
Technical details of the integration algorithm can be found
elsewhere.\cite{hoye,JCP_2007_126_244510} The integration steps and length of the
simulations were chosen after performing a number of tests.

The GDI was divided in two parts. In the first part we perform an integration
following the scheme given in Eq.~(\ref{gdi1}). The first point was chosen to be
$T^*=(\beta\epsilon)^{-1}=0.35$, $\mu/\epsilon=12$. The integration was carried out using a temperature
step of $\Delta T^*=0.025$. At each step we run long simulations ($2\times 10^6$
cycles, each cycle including $M/3$ insertion/deletion attempts; averages are
taken over the second half of the simulation). The integration was carried out
up to $T^*=1.25$ (or $\beta\epsilon=0.80$), where we get coexistence for
$\mu/\epsilon = 11.947 \pm 0.001$.

The second part of the integration was carried out using the scheme given in
Eq.~(\ref{gdi2}). The starting point was that defined by the previous integration
($T^*=1.25$, $\beta\epsilon=0.80$,  $\beta \mu = 9.558$). The integration step was then
$\Delta (\beta \mu) = -0.020$, and the length of the  simulations was about
$1 \times 10^6$ cycles (for each system and step, averaging over the second half
of each run). As in the previous part, a number of additional simulations with
larger integration steps, smaller system sizes, and shorter runs, were carried
out in order to test the integration accuracy and to estimate error bars.

Simulations were launched to execute 201 integration steps (to reach, in principle,
a final value of $\beta \mu = 5.558$). We observed that for this line GDI required
both large systems and precise estimates of the integrand in order to avoid the
collapse of the method before reaching the F--T4 discontinuous equilibrium found
with WL simulations. For instance, using large simulations (precise integrands)
with $L=60$, $\Delta (\beta \mu) = -0.05$, it was possible to get good estimates
both for the triple point T4--F--T3 and for the temperature at which the density of
phases F and T3 are equal at equilibrium, but shortly after reaching the latter point
the algorithm failed due to the transition of one of the phases into the other. With
$L=120$ the integration stayed stable until reaching the expected final value of
$\beta \mu\simeq 5.56$, which was found to be consistent, within statistical
uncertainty, with the value of the F--T4 equilibrium obtained
through WL simulation at $\beta\epsilon=0.63$.

From the results obtained with GDI, together with those previously reported on
the F--T3 transition, we can estimate the position of two of the {\em special}
points in the phase diagram, namely the triple point T4--F--T3 and the point of
maximum temperature for the F--T4 equilibrium (at this point both phases have the
same density). These results, together with those of the change of the transition
order of the F--T4 equilibrium, are collected in Table~\ref{table-specials}.

\begin{table}
\begin{center}
\begin{ruledtabular}
\begin{tabular} {c|ccc}
    Point        & TP (T4-F-T3) & HT (F-T4)  & DC (F-T4)      \\
\hline
   $\beta\epsilon$    &  0.660(2)\phantom{0} &  0.619(1)  &  $\sim$ 0.75\phantom{0} \\
  $\beta \mu$ &  7.78(2)\phantom{00} &   6.35(2)\phantom{0}  &  $\sim$ 4.0\phantom{0} \\
 $\eta_{F}$   &  0.892(5)\phantom{0} &   0.745(1) &  $\sim$ 0.69 \\
 $\eta_{T3}$  &  0.892(5)\phantom{0} &  ---       &  --- \\
 $\eta_{T4}$  &  0.7491(1)   &  0.745(1)  &  $\sim$ 0.69
\end{tabular}
\end{ruledtabular}
\caption{Singular points of the phase diagram: TP stands for triple point,
HT for high temperature end-point, and DC for the change from discontinuous to continuous
behavior of the fluid-T4 transition.
Error bars are estimated
by comparing results from different GDI trajectories.}
\label{table-specials}
\end{center}
\end{table}

\subsection{LFMT calculations}

We shall now apply the density functional obtained in Sec.~\ref{sec:LFMT}
to determine the temperature-density phase diagram of this fluid. To this purpose
we have to study the three phases involved: the uniform fluid and the two
solid lattices (T3 and T4, c.f.~Fig.~\ref{fig:t3}).

\subsubsection{Uniform fluid}

If every site has the same average occupancy $\rho$ (hence a packing
fraction $\eta=3\rho$),
the free energy per unit volume (in $k_BT$ units) will be
\begin{equation}
\begin{split}
\Phi =&\, \rho\ln\rho+(1-\rho)\ln(1-\rho)-4(1-3\rho)\ln(1-3\rho) \\
&+ 3(1-4\rho)\ln(1-4\rho+\rho_{14})+6\rho\ln\left(1-\frac{\rho_{14}}{\rho}\right),
\end{split}
\end{equation}
where $\rho_{14}=\rho_{14}(\rho,\rho,\rho,\rho)$ is
\begin{equation}
\begin{split}
\rho_{14} =&\, \frac{1}{2(1-\kappa)}\Big\{2(2-\kappa)\rho-1 \\
& \left. +\sqrt{1-4(2-\kappa)\rho+ 4(4-3\kappa)\rho^2}\right\}.
\end{split}
\end{equation}
The pressure can then be obtained as
\begin{equation}
\beta p =
\rho^2\frac{\partial(\Phi/\rho)}{\partial\rho}=
\ln\left[\frac{(1-3\rho)^4}{(1-4\rho+\rho_{14})^3(1-\rho)}\right],
\end{equation}
and the chemical potential is given by $\beta\mu=(\Phi+\beta p)/\rho$.

\subsubsection{T3 solid phase}

The T3 solid occupies one of the three sublattice shown in Fig.~\ref{fig:t3}(a).
From this figure it is clear that the density profile will take one out of
two values: The sites of the occupied sublattice ---marked with large circles
in Fig.~\ref{fig:t3}(a)--- will have a higher density $\rho_A$, whereas 
the remaining ones will have a lower density $\rho_B$ (by symmetry
this value is the same for all those sites). The packing fraction will then
be $\eta = \rho_A+2\rho_B$.

If we regard the lattice as a tiling of rhombii,
we can realize that there are three different kinds of them: Those with
$\rho_A$ at positions $1$ and $4$ and $\rho_B$ at positions $2$ and $3$,
those with $\rho_A$ only at position $2$, and 
those with $\rho_A$ only at position $3$. There is the same number of each
kind. The contribution of the two latter to the rhombic cavities will be
the same, but different from the contribution of the former. Thus
\begin{equation}
\sum_{C\in\mathcal{W}_4}\frac{\Phi_0(C)}{V}=\Phi_0(\rho_A,\rho_B,\rho_B,\rho_A)+
2\Phi_0(\rho_B,\rho_A,\rho_B,\rho_B),
\end{equation}
with $\Phi_0(\boldsymbol\rho)$ given by (\ref{eq:Phi0}). As for the
triangular cavities, $\sum_i\rho_i=\rho_A+2\rho_B=\eta$ for any of them, so
\begin{equation}
\sum_{C\in\mathcal{W}_3}\frac{\Phi_0(C)}{V}=2\eta+2(1-\eta)\ln(1-\eta).
\end{equation}
Dimers do not contribute to the functional, so the last contribution will be
\begin{equation}
\begin{split}
\sum_{C\in\mathcal{W}_1}\frac{\Phi_0(C)}{V}=&\,\frac{\eta}{3}+
\frac{1}{3}(1-\rho_A)\ln(1-\rho_A) \\
&+\frac{2}{3}(1-\rho_B)\ln(1-\rho_B).
\end{split}
\end{equation}

Putting all together and adding the ideal part the result is
\begin{widetext}
\begin{equation}
\begin{split}
\Phi(\rho_B;\eta) =&\,
\frac{1}{3}(1-\eta+2\rho_B)\ln(1-\eta+2\rho_B)+\frac{2}{3}(1-\rho_B)\ln(1-\rho_B)-4(1-\eta)\ln(1-\eta)\\
&+(1-2\eta+2\rho_B)\ln(1-2\eta+2\rho_B+\xi_1)+2(1-\eta-\rho_B)\ln(1-\eta-\rho_B+\xi_2)\\
&+(\eta-2\rho_B)\left\{2\ln(\eta-2\rho_B-\xi_1)-\frac{5}{3}\ln(\eta-2\rho_B)\right\}
+2\rho_B\left\{2\ln(\rho_B-\xi_2)-\frac{5}{3}\ln\rho_B\right\},
\label{eq:PhiT3}
\end{split}
\end{equation}
\end{widetext}
where we have eliminated $\rho_A=\eta-2\rho_B$, and
\begin{eqnarray}
\xi_1 &=& \rho_{14}(\rho_A,\rho_B,\rho_B,\rho_A), \\
\xi_2 &=& \rho_{14}(\rho_B,\rho_A,\rho_B,\rho_B).
\end{eqnarray}

The free energy as a function of $\eta$ is obtained by minimizing the function (\ref{eq:PhiT3})
with respect to $\rho_B$ (always taking the solution $\rho_B\le\eta/3$). For $\kappa>\kappa_t
=0.5086\dots$ (see below) there is a first order transition from a homogeneous fluid to
the T3 solid [see Fig.~\ref{T-eta-fmt}]. In the limit $\kappa=1$ where the model becomes
identical to the hard hexagon model, we find a wide first order transition.
This model has been previously solved within the LFMT approach in
Ref.~\onlinecite{lafuente:2003b}. LFMT is a mean-field-like theory, and therefore
all second order transitions have a parabolic behavior of the order parameter
---corresponding to a critical exponent $\beta=1/2$. The exponent of the hard hexagon
model (like that of the 3-state Potts model) is\cite{JPAMG_1980_13_L61,baxter2}
$\beta=1/9$. Looking at Figure~7 of Ref.~\onlinecite{lafuente:2003b} one must
admit that a discontinuous function is a better approximation to the behavior
of the order parameter than a parabolic, mean-field one. So although not quite
satisfying, the result is quantitatively not too inaccurate. This also reflects
in the fact that the pressure and chemical potential at the transition is rather
close to the exact value, and it remains so for all $\kappa_t<\kappa<1$, as
Figs.~\ref{bP-T} and \ref{bmu-T} show.

\subsubsection{T4 solid phase}

\begin{figure}
\includegraphics[width=80mm,clip=]{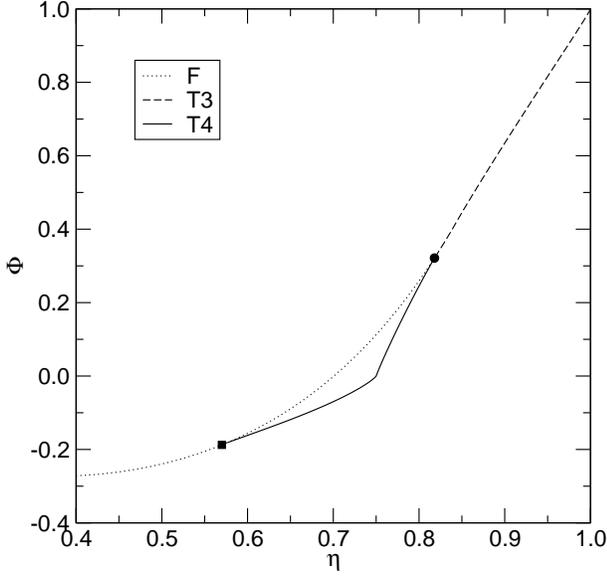}
\caption{Free energy (in $k_BT$ units) per unit volume as a function of
the packing fraction $\eta$, for $\kappa=e^{-\beta\epsilon}=0.3679$, below
the triple point. Dotted line represents the free energy of the uniform fluid;
solid line that of the T4 phase, and dashed line that of the T3 phase. The
filled square represents the bifurcation point F--T3, and the filled circle
the bifurcation point T3--T4.
Notice the discontinuity of the derivative of the free energy at $\eta=0.75$,
the close packing of the NNN exclusion lattice gas. For $\eta>0.75$ the
free energy of the T4 phase is concave, so T4--T3 coexistence always occurs
with a T4 phase at $\eta=0.75$, and both the chemical potential and the
pressure jump discontinuously at this transition.}
\label{energy}
\end{figure}

The T4 solid occupies one of the four sublattice shown in Fig.~\ref{fig:t3}(b).
Again the density profile will take either the value $\rho_A$ at the sites of the
occupied sublattice ---marked with large circles in Fig.~\ref{fig:t3}(b)---
or the value $\rho_B$ at the remaining sites (the same for all of them,
by symmetry). The packing fraction will now be $\eta = \rho_A
+3\rho_B$.

As a tiling of rhombii the lattice contains four kinds of them,
each with an A site at one of the four positions. There is the same amount of
each type. Thus
\begin{equation}
\begin{split}
\sum_{C\in\mathcal{W}_4}\frac{\Phi_0(C)}{V}=&\,
\frac{3}{2}\Phi_0(\rho_A,\rho_B,\rho_B,\rho_B) \\
&+\frac{3}{2}\Phi_0(\rho_B,\rho_A,\rho_B,\rho_B).
\end{split}
\end{equation}
As for the triangles, one fourth of them have an A site and two B sites, and
three fourths have three B sites, so
\begin{equation}
\begin{split}
\sum_{C\in\mathcal{W}_3}\frac{\Phi_0(C)}{V}=&\,\frac{\rho_A+2\rho_B}{2}+
\frac{1}{2}(1-\rho_A-2\rho_B) \\
&\times \ln(1-\rho_A-2\rho_B)+\frac{3}{2}\rho_B \\
&+\frac{3}{2}(1-3\rho_B)\ln(1-3\rho_B).
\end{split}
\end{equation}

Finally, the contribution of the point-like cavities (those of $\mathcal{W}_1$)
is
\begin{equation}
\begin{split}
\sum_{C\in\mathcal{W}_1}\frac{\Phi_0(C)}{V}=&\,\frac{\eta}{4}+
\frac{1}{4}(1-\rho_A)\ln(1-\rho_A) \\
&+\frac{3}{4}(1-\rho_B)\ln(1-\rho_B),
\end{split}
\end{equation}
because $1/4$ of the sites are of type A and $3/4$ of type B.

Putting all together and adding the ideal part the result is
\begin{widetext}
\begin{equation}
\begin{split}
\Phi(\rho_B;\eta) &= \frac{3-4\eta}{2}\left\{\ln\left(1-\frac{4\eta}{3}+\lambda_1\right)+
\ln\left(1-\frac{4\eta}{3}+\lambda_2\right)\right\}+\frac{4\eta-9\rho_B}{2}\ln\left(\frac{4\eta}{3}-3\rho_B-
\lambda_1\right)\\
&+\frac{3\rho_B}{2}\{\ln(\rho_B-\lambda_1)+2\ln(\rho_B-\lambda_2)\}+\frac{3}{4}(1-\rho_B)\ln(1-\rho_B)-
(1-3\rho_B)\ln(1-3\rho_B)\\
&-\frac{15}{4}\rho_B\ln\rho_B-3\left(1-\frac{4\eta}{3}+\rho_B\right)\ln\left(1-\frac{4\eta}{3}+\rho_B\right)
+\frac{1}{4}\left(1-\frac{4\eta}{3}+3\rho_B\right)\ln\left(1-\frac{4\eta}{3}+3\rho_B\right)\\
&-\frac{5}{4}\left(\frac{4\eta}{3}-3\rho_B\right)\ln\left(\frac{4\eta}{3}-3\rho_B\right),
\end{split}
\end{equation}
\end{widetext}
where we have eliminated $\rho_A=\eta-3\rho_B$, and
\begin{eqnarray}
\lambda_1 &=& \rho_{14}(\rho_A,\rho_B,\rho_B,\rho_B), \\
\lambda_2 &=& \rho_{14}(\rho_B,\rho_A,\rho_B,\rho_B).
\end{eqnarray}

As for T3, the free energy of the equilibrium phase is obtained by minimization
of this function with respect to $\rho_B$ (always choosing the solution
$\rho_B\le\eta/4$). In the limit $\kappa=0$ the model is equivalent to a
lattice gas with NNN exclusion. In this limit we obtain a wide first order
transition from a uniform fluid to a T4 solid, which again is found to be
continuous in the simulations. The values of the pressure and chemical
potential for this transition are nevertheless rather accurately predicted
(see Figs.~\ref{bP-T} and \ref{bmu-T}), so the same considerations as for
the F--T3 transition in the hard hexagon ($\kappa=1$) limit hold here.
We find a first-order F--T4 transition all the way up to $\kappa_c=0.5403\dots$,
where it coalesces to an end-point (at $\eta_c=0.7405\dots$).
It is to be noticed that this point is obtained with high accuracy (simulations
yield $\kappa_c\approx 0.538$ and $\eta_c\approx 0.745$; see
Table~\ref{table-specials}), and that simulations also find a
first-order F--T4 transition near this point (see Fig.~\ref{T-eta}).

When $\kappa=0$ the model has a close-packing at $\eta=3/4$; however, for
any $\kappa>0$ this limit can be crossed, although at a very high energetic
cost. Once this cost is paid, the system greatly diminishes its entropy by
reordering itself in a T3 structure. Hence the transition that is found,
for all $\kappa$ up to not too far from $\kappa_t$, between a close-packed
T4 solid and a nearly closed-packed T3 solid, which is also observed in
the simulations (see Fig.~\ref{T-eta}). What happens with the free energy
at this T4--T3 transition is very peculiar and it is illustrated in
Fig.~\ref{energy}. The derivative of the free energy per unit volume
(in $k_BT$ units) with
respect to the packing fraction is discontinuous at $\eta=3/4$. The
free energy of the T4 phase is concave beyond this point, so there is
a T4--T3 coexistence, but it does not satisfy the standard conditions
of phase equilibria. As a matter of fact, both the pressure and the
chemical potential jump discontinuously at this point. This creates
a ``forbidden'' region in the pressure-temperature and the chemical
potential-temperature, which can be observed in Figs.~\ref{bP-T} and
\ref{bmu-T}.

At the value $\kappa_t$ there is a F--T3--T4 triple point, which the
theory predicts very close to the value obtained in the simulations,
$\kappa_t\approx 0.517$ (see Table~\ref{table-specials}).

In the range $\kappa_t<\kappa<\kappa_c$ a reentrant T4--F transition is
found before the F--T3 transition occurs. This reentrant behavior also appears
in the simulations, although the coexistence region is wider because the
F--T3 transition is continuous (see Fig.~\ref{T-eta}).

\section{Discussion}

\begin{figure}
\includegraphics[width=80mm,clip=]{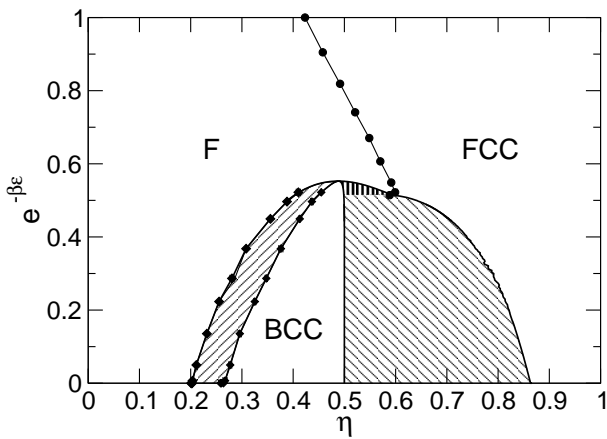}
\caption{Temperature-density phase diagram for the 3d counterpart of
the model discussed in this article, on a simple cubic lattice.}
\label{fig3d}
\end{figure}

From Figs.~\ref{T-eta}, \ref{bP-T}, and \ref{bmu-T}, we see that the
theoretical results for
the equilibrium between the ordered phases are in reasonably good agreement
with the simulation. The theoretical estimations of the triple point and
the high temperature end-point of the F--T4 equilibrium are also well
described. On the other hand the LFMT description of the order-disorder
transitions, both at low and high temperature is less accurate. The theory
predicts first order transitions whereas they are continuous in both limits.
We have argued that this is so because the behavior of the order parameter
is very sharp (the critical exponent $\beta=1/9$), so much that a discontinuous
function is a better approximation to it than the simple parabolic 
behavior predicted by any mean-field-like theory (like this one). The theory
can in principle be refined by considering larger ``maximal cavities''. Although
it would produce a far more complicated theory than the one presented here,
and it is expected that its accuracy would increase, 
it is doubtful that the order of the transition would be corrected. No
matter how much we complicate the theory, it does not cure its mean-field
behavior near the transitions. This is also the reason why first-order
transitions are described much better.

In favor of this argument is the fact that, overall, the agreement
between simulation and theoretical transition lines in
the planes $\beta\mu$--$T$ and $\beta p$--$T$ is rather good.

When comparing the phase diagram for the two-dimensional system on the
triangular lattice with that of the three-dimensional system\cite{hoye}
[c.f.~Fig.~\ref{fig3d}] in a simple cubic lattice
we find a couple of qualitative differences.
The first one concerns the nature of the order-disorder transition
at low temperature, which is continuous in two dimensions and discontinuous
in three dimensions. This difference can be understood in terms of
the different dimensionality, and can be related with the critical
behavior of Potts models \cite{RMP_1982_54_235} in two and three dimensions.

The second relevant difference arises when comparing the transitions
between the two ordered phases. In two dimensions, at solid-solid equilibrium
the high density solid is nearly close packed for a wide range of
temperatures, whereas this is not the case for its three dimensional
counterpart even at the lowest temperatures. 
This difference can be explained as follows. At low temperature 
phase equilibrium is essentially controlled by the condition of
minimum energy. The energy per unit volume of the closed packed
configuration in two dimensions is $\bar u^*=\mathcal{U}/M\epsilon=3$.
At slighly lower $\eta$ there are vacancies. The way in which they
minimize the energy is by not being nearest neighbors in the solid
lattice (i.e.\ next-nearest neighbors in the underlying lattice).
This way each vacancy reduces the energy by $6\epsilon$. Thus, for
$2/3<\eta\le 1$, $\bar u^*(\eta)=2\eta-1$. If we consider, at the
same density, a system separated into a close-packed T3 and a T4
phase, then $N_{\rm T3}+N_{\rm T4}=N$ and $3N_{\rm T3}+4N_{\rm T4}=M$,
and the energy of this system will be $U=3\epsilon N_{\rm T3}$.
Hence $\bar u^*(\eta)=4\eta-3$. Since $4\eta-3<2\eta-1$ for all
$\eta<1$, then the phase separated system is energetically favored.
The same argument for the three-dimensional model of H\o ye \emph{et al.}\
\cite{hoye} yields the same energy, $\bar u^*(\eta)=6\eta-3$
(for $3/4 \le \eta \le  1)$ in both cases, so
in the three dimensional system
the entropy does play a role in defining the density of the high
density solid, and the number of
vacancies does not go to zero when approaching $T=0$.

The similarity between the phase diagrams in two and three dimensions
is remarkable, though. Peculiar features like the reentrant fluid phase
or the vertical line at the closest packing of the loose solid when
it coexists with the dense one, appear in both cases.

Much to our surprise, we have realized that the LFMT for the three-dimensional
model is far more complicated than that of the two-dimensional one. The
reason is that the soft repulsion at NNN allows for maximal cavities with
up to four particles at the same time. This not only introduces a much
larger set of cavities to elaborate the density functional, but also
the corresponding expressions for the $\Phi_0$ functions is very 
cumbersome and hard to handle. On the other hand, given the similarity
between the phase diagrams, it seems that the physics of the model is
already well captured by the two-dimensional version.

\acknowledgements

We acknowledge support from Direcci\'on General de
Investigaci\'on Cient\'{\i}fica y T\'ecnica under projects no.\
MAT2007-65711-C04-04 (N.G.A.\ and E.L.) and MOSAICO (J.A.C.\ and J.A.C.),
and the Direcci\'on General de Universidades
e Investigaci\'on de la Comunidad de Madrid under project MOSSNOHO-CM
(S0505/ESP/0299). J.\ A.\ Capit\'an is supported by
a contract from Comunidad de Madrid and Fondo Social Europeo.


\end{document}